\documentclass{osa-article}

\journal{osajournal}
\usepackage{multirow}
\usepackage{bm}
\articletype{letter}
\begin{document}
\title{Investigation of the electro-optic effect in high-$Q$ 4H-SiC microresonators}

\author{Ruixuan Wang,\authormark{1}, Jingwei Li \authormark{1}, Lutong Cai \authormark{1}, and Qing Li\authormark{*1}}

\address{\authormark{1}Department of Electrical and Computer Engineering, Carnegie Mellon University, Pittsburgh, PA 15213, USA}

\email{\authormark{*}qingli2@andrew.cmu.edu} %% email address is required



\begin{abstract}
Silicon carbide (SiC) recently emerged as a promising photonic and quantum material owing to its unique material properties. Here, we carry out an exploratory investigation of the electro-optic effect in high-quality-factor 4H-SiC microresonators. Our findings confirm the existence of the Pockels effect in 4H-SiC for the first time. The extracted Pockels coefficients show certain variations among 4H-SiC wafers from different manufacturers, with the magnitudes of $r_{13}$ 
 and $r_{33}$ estimated to be in the range of (0.3-0.7) pm/V and (0-0.03) pm/V, respectively.   
\end{abstract}
\noindent 
Silicon carbide (SiC), a wide-bandgap compound semiconductor composed of silicon and carbon, is playing an increasingly important role in modern power electronics including electric-vehicle charging \cite{Kimoto_SiC_review}. In addition, considerable research efforts have been devoted to studying its potential for photonic and quantum applications, resulting in the demonstration of low-loss SiC-on-insulator (SiCOI) integrated photonics platforms and the discovery of a variety of color centers with promising quantum properties \cite{Awschalom_SiC_qubit, Vuckovic_SiC_review}. Among its many polytypes such as 3C, 4H and 6H, 4H-SiC gains the most attention from the industry as single-crystal 4H-SiC substrates up to six inches are already commercially available. Thanks to the continuously improved material quality, 4H-SiCOI has achieved a much lower propagation loss compared to other polytypes \cite{Ou_4HSiC_combQ}, which is a key enabling factor for applications such as low-power optical parametric oscillation \cite{Vuckovic_4HSiC_soliton} and broadband microcomb generation \cite{Li_4HSiC_comb}.

For integrated photonic circuits, one important functionality that underpins applications including optical signal modulation and switching is the electro-optic (EO) tuning \cite{Soref_EO}. That is, the application of an external electric field in the radio frequency or below leads to a change in the optical absorption and/or the refractive index of the hosting material. Two prominent examples are the plasma dispersion effect in silicon \cite{Lipson_Si_modulator} and the Pockels effect in lithium niobate \cite{Loncar_LN_modulator}. In the SiCOI integrated photonics platform, only 3C-SiC has been experimentally shown to possess the fast Pockels effect and hence is capable of delivering modulation bandwidths on the order of gigahertz (GHz) \cite{Adibi_SiC_EO, Loncar_SiC_modulator}. While theoretical calculations predict a similar Pockels effect for 4H-SiC \cite{SiC_EO_theory}, to date only slow EO modulation based on carrier depletion was reported \cite{4H-SiC-EO-carrier}. As such, it is unclear whether GHz-level EO modulation/switching is attainable in 4H-SiC \cite{Ou_SiC_review}.

\begin{figure}[ht]
\centering
\includegraphics[width=0.75\linewidth]{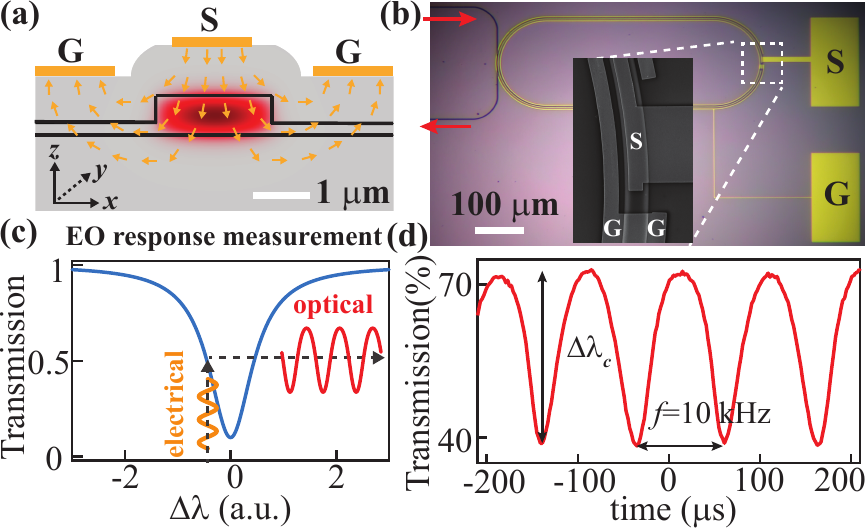}
\caption{(a) Cross-sectional view of a 4H-SiC rib waveguide with the signal (S) and ground (G) electrodes implemented near the waveguide mode (red colored). The 4H-SiC waveguide is surrounded by PECVD oxide. (b) Optical micrograph of a SiC racetrack resonator along with the electrical pads. The inset shows the scanning electron micrograph of the conjunction between the signal electrode on top of the waveguide and the connecting wire to the pad. (c) The electro-optic effect is quantified by measuring the optical transmission of a high-$Q$ SiC microresonator  in response to an applied electrical signal. (d) Exemplary transmission measured in a high-$Q$ racetrack resonator with a sinusoidal electrical input at a frequency of 10 kHz. The corresponding resonance shift ($\Delta \lambda_c$) is inferred from the modulation amplitude.}
\label{Fig_Schematic}
\end{figure}

In this Letter, we carry out a first experimental investigation of the Pockels effect in 4H-SiC. For this purpose, low-loss microresonators are fabricated on on-axis (i.e., $c$-axis along $z$ in Fig.~\ref{Fig_Schematic}(a)), semi-insulating 4H-SiCOI wafers, which are obtained using a customized bonding and polishing process (NGK Insulators) \cite{Li_4HSiC_comb}. One such example is provided in Fig.~\ref{Fig_Schematic}(b), where light from a tunable laser source is coupled to a racetrack resonator using on-chip waveguides and grating couplers \cite{Li_4HSiC_comb}. To realize efficient EO tuning, electrodes are placed near the waveguide mode without introducing additional optical losses (Fig.~\ref{Fig_Schematic}(a)). Given the large resistivities from the 4H-SiC and surrounding oxide layers, there is no current flow between the signal and ground electrodes. (The resistance between the signal and ground pads is experimentally verified to be larger than 10 M$\Omega$.) Instead, the EO effect is expected to be dominated by the field-induced change in the refractive index of SiC (i.e., the Pockels effect), which in turn shifts the cavity's resonance wavelength. 

As illustrated in Fig.~\ref{Fig_Schematic}(c), we quantify the EO effect in SiC microresonators by fixing the input laser's wavelength at a proper bias and applying a sinusoidal electrical signal with varied amplitudes and frequencies. If the EO-induced wavelength shift ($\Delta \lambda_c$, see Fig.~1(d)) is smaller than the cavity linewidth, we typically set the bias corresponding to the mid-point of the optical transmission (i.e., detuning equals half of the linewidth). For larger wavelength shifts, however, the bias is chosen at the bottom of the transmission to increase the measurement range (the maximum measurable $\Delta \lambda_c$ is approximately two to three times of the cavity linewidth). With the information of the bias, $\Delta \lambda_c$ is then inferred from the modulation amplitude in the laser transmission, which should have the same frequency as the electrical signal (see one example in Fig.~\ref{Fig_Schematic}(d)). In this regard, a high-$Q$ resonance is beneficial given that the minimally detectable $\Delta \lambda_c$ is proportional to the cavity's linewidth.  

\begin{figure}[ht]
\centering
\includegraphics[width=0.75\linewidth]{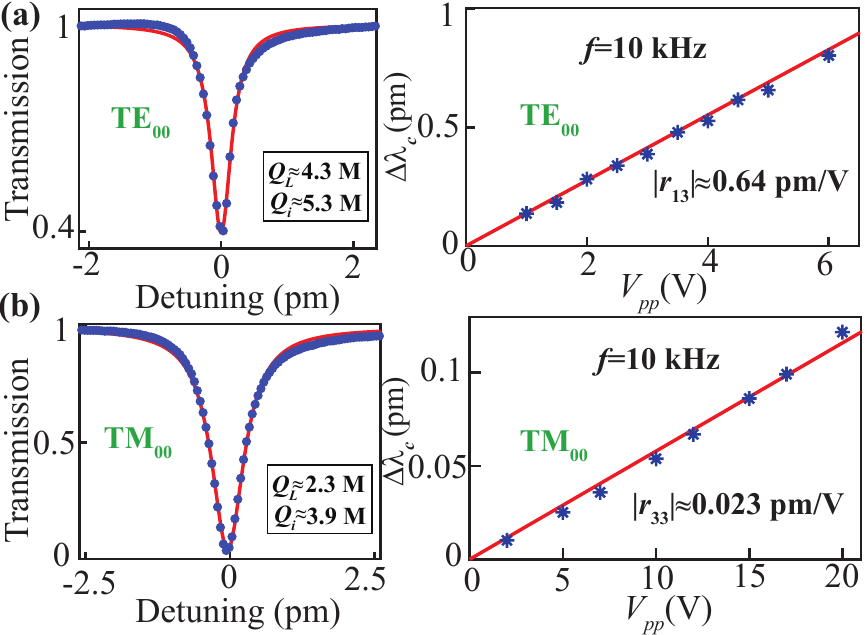}
\caption{The left figure in (a) and (b) plots the swept-wavelength transmission of the select high-$Q$ resonance for the fundamental TE (TE$_{00}$) and TM (TM$_{00}$) modes in 4H-SiC (Norstel) racetrack resonators in the 1550 nm band. $Q_L$ and $Q_i$ are the loaded $Q$ and intrinsic $Q$, respectively. The figure on the right displays the measured resonance wavelength shifts (blue stars) and the corresponding linear fit (red solid line) for varied peak-to-peak voltages ($V_{pp}$) of a sinusoidal electrical signal at a fixed frequency of 10 kHz.}
\label{Fig_EO}
\end{figure}

In Fig.~\ref{Fig_EO}, measurement results for two racetrack resonators fabricated on a 4H-SiC wafer from ST Microelectronics (formerly known as Norstel AB and hereinafter referred to as "Norstel" for short) are provided. Both devices have an approximate waveguide width of 2500 nm, a nominal SiC thickness of 850 nm, and a pedestal layer of 250 nm. The bending radius of the racetrack resonator is chosen to be 100 $\mu$m (with circumference near 1.3 mm) to minimize the radiation loss. After ebeam lithography and dry etching, a 1-$\mu$m-thick PECVD oxide is deposited on top of the SiC microresonator, and electrodes made of Ti/Au layers (30 nm Ti and 70 nm Au) are fabricated using the ebeam evaporation and liftoff process. To avoid the optical loss from the metal layer, the ground electrodes are placed 1 $\mu$m away from the sidewall of the SiC waveguides.

As can be seen from Fig.~\ref{Fig_EO}, optical resonances with intrinsic $Q$s up to 5 million are observed for the transverse-electric (TE, dominant electric field in the $x-y$ plane as indicated in Fig.~\ref{Fig_Schematic}(a)) and transverse-magnetic (TM, dominant electric field along the $z$ axis as indicated in Fig.~\ref{Fig_Schematic}(a)) resonant modes in the 1550 nm wavelength band. These optical $Q$s are comparable to the state of the art \cite{Ou_4HSiC_combQ, Vuckovic_4HSiC_soliton}, which make our EO measurement extremely sensitive to even a small amount of resonance shift. In addition, by focusing on the amplitude modulation at the exact same frequency of the applied electrical signal, we can suppress the ubiquitous thermal noise, which causes random fluctuations (albeit small) in the bias point due to wavelength instabilities from both the microresonator and the input laser. 

\begin{figure}[ht]
\centering
\includegraphics[width=0.75\linewidth]{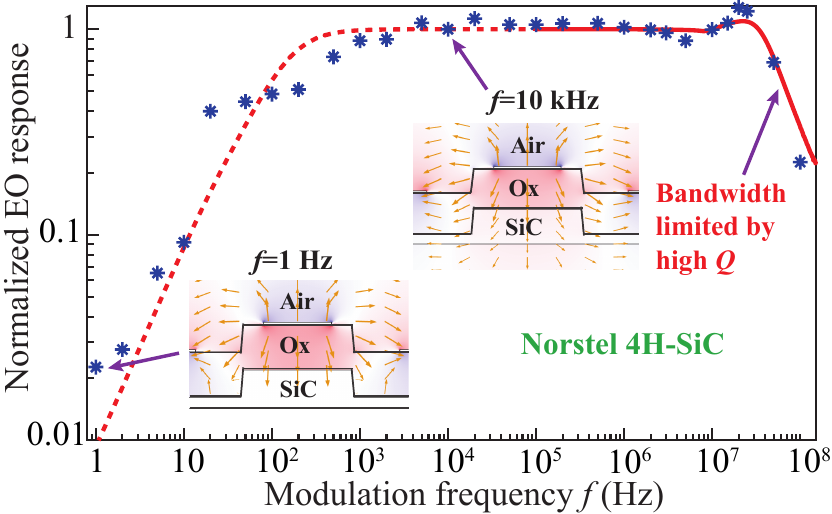}
\caption{The blue stars are the measured and normalized (using $f=10$ kHz as the reference) modulation amplitude of the TE$_{00}$ mode in Fig.~\ref{Fig_EO}(a) at varied testing frequencies for a fixed $V_{pp}=2$ V. The red dashed line is the simulated EO response using SiC's actual resistivity of $1.7 \times 10^{9}\ \Omega \cdot \text{cm}$, while the red solid line depicts the expected EO modulation amplitude in an optical microcavity with a linewidth of 45 MHz (corresponding loaded $Q$ of 4.3 million). The two insets plot the simulated electric field at $f=1\ $Hz and $f=10\ $kHz.}
\label{Fig_freq}
\end{figure}

To understand the significant difference in the EO responses between the TE and TM polarizations and extract their corresponding EO coefficients, we utilize the fact that on-axis 4H-SiC wafers have a hexagonal crystal structure ($C_{6v}$ point group) with an extraordinary refractive index ($n_e$) along the $z$ axis and an ordinary refractive index ($n_o$) in the $x-y$ plane (see Fig.~1(a)). For the electrode configuration shown in Fig.~\ref{Fig_Schematic}(a), the dominant component of the electrical signal is along the $z$ axis (denoted as $\tilde{\bm E}_z$). A straightforward analysis based on the electro-optic tensor yields the change of the refractive index as $\Delta n_\text{SiC, TE}=-n_o^3r_{13} \tilde{\bm E}_z/2$ and $\Delta n_\text{SiC, TM}=-n_e^3r_{33} \tilde{\bm E}_z/2$, with $r_{13}$ and $r_{33}$ being the EO coefficients of 4H-SiC. The corresponding resonance wavelength shift ($\Delta \lambda_0$) can be computed using the following relationship \cite{Lan_index_sensing}:
\begin{equation}
\frac{\Delta \lambda_0}{\lambda_0}= \frac{\iint_\text{core}\epsilon_r  |\bm E(x,z)|^2dxdz}{\iint \epsilon_r |\bm E (x,z)|^2 dxdz}\cdot \frac{<\Delta n_\text{SiC}>}{n_\text{SiC}}=\Gamma\frac{<\Delta n_\text{SiC}>}{n_\text{SiC}}, \label{Eq1}
\end{equation}
where $\epsilon_r$ is the relative permittivity; $\bm E(x,z)$ is the electric field of the optical mode; $\Gamma$, which is the mode confinement factor, represents the fractional optical energy stored within the waveguide core ($\Gamma \approx 1$ for the waveguide modes studied in this work); and $<\Delta n_\text{SiC}>$ is the spatially averaged change in the refractive index of SiC due to the EO effect. Considering that our EO measurement only determines the magnitude of $\Delta \lambda_0$ (which is denoted as $\Delta \lambda_c$ in Fig.~\ref{Fig_Schematic}(d) for distinction), we obtain
\begin{align}
&\Delta \lambda_c (\text{TE})=\frac{\lambda_0 n_o^2\Gamma}{2}|r_{13}<\tilde{\bm E}_z>|, \label{Eq2}\\
&\Delta \lambda_c (\text{TM})=\frac{\lambda_0 n_e^2\Gamma}{2}|r_{33}<\tilde{\bm E}_z>|, \label{Eq3}
\end{align}
where $<\tilde{\bm E}_z>$ is the spatially averaged electrical signal inside the SiC waveguide. By numerically simulating the electric field for a given voltage (see examples in Fig.~\ref{Fig_freq}), we extract the magnitude of $r_{13}$ and $r_{33}$ for the Norstel 4H-SiC to be around 0.64 pm/V and 0.023 pm/V, respectively. 

In order to investigate the nature of the observed EO effect, we vary the frequency of the electrical signal for the TE mode in Fig.~\ref{Fig_EO}(a) with a fixed voltage and bias, and normalize the measured modulation amplitude to that of $f=10$ kHz. As observed in Fig.~\ref{Fig_freq}, the EO response becomes significantly smaller at modulation frequencies below 100 Hz, and nearly disappears at 1 Hz or below \cite{Loncar_SiC_modulator}. In fact, the inferred $\Delta\lambda_c$ at DC is still much smaller than 1 pm even with 100 V applied voltage. While surprising for the supposedly high-quality semi-insulating 4H-SiC wafers, this kind of frequency dependence can be explained by the significantly smaller resistivity of the 4H-SiC layer compared to that of the surrounding oxide. For example, the average resistivity of the Norstel 4H-SiC wafer used in this work is measured around $1.7\times 10^9\ \Omega \cdot \text{cm}$ \cite{Li_Kerr_compare}, which is 7-8 orders of magnitude lower than that of silicon dioxide. The two simulation examples for $f=1$ Hz and $f=10$ kHz included in Fig.~\ref{Fig_freq} suggest that the degraded EO response at low modulation frequencies can be primarily attributed to a smaller electric field within the SiC waveguide (i.e., $|<\tilde{\bm E}_z>|$ in Eq.~\ref{Eq2}), which leads to a reduced $\Delta\lambda_c$. A more intuitive explanation is that near the DC frequency, the majority of the applied voltage falls on the oxide layer instead of the SiC waveguide. Nevertheless, our simulation predicts that the impact of the limited resistivity from 4H-SiC should become negligible for high-enough modulation frequencies, hence acting like a high-pass filter for the applied electrical signal (red dashed line in Fig.~\ref{Fig_freq}). This frequency behavior is confirmed by the experimental data, which is also the main reason for us to characterize the EO coefficients at $f=10$ kHz in Fig.~\ref{Fig_EO}. On the high-frequency side, the modulation amplitude maintains a flat shape until the testing frequency gets close the cavity linewidth. Notably, we see a slight increase in the EO response when the modulation frequency equals the bias detuning (set at half of the linewidth), which is followed by a characteristic roll-off when the modulation frequency exceeds the cavity linewidth (45 MHz for the example studied in Fig.~\ref{Fig_freq}). Unlike the low-frequency case, such frequency behaviors can be well explained by assuming a constant $\Delta\lambda_c$ (red solid line in Fig.~\ref{Fig_freq}), indicating that our EO modulation bandwidth is simply limited by the narrow linewidth of the optical resonance. One can intuitively understand the roll-off in the high-frequency range as when the modulation frequency gets close to the cavity linewidth or above, the generated optical sidebands become out of resonance, hence resulting in diminished EO responses. 

\begin{figure}[ht]
\centering
\includegraphics[width=0.75\linewidth]{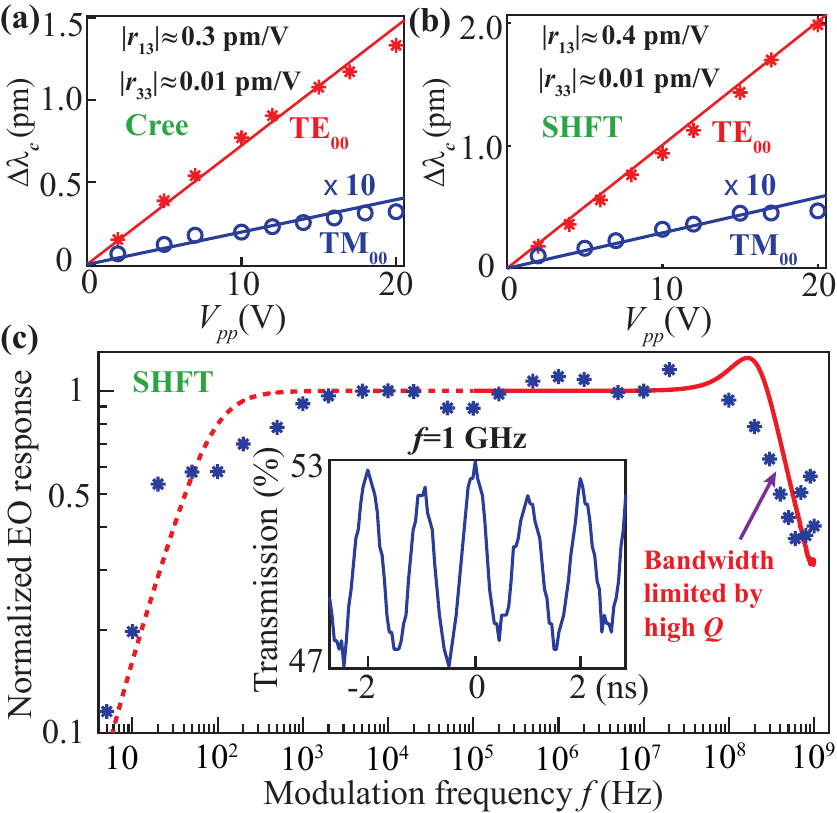}
\caption{(a) Measured EO-induced wavelength shifts (markers) and the corresponding linear fitting (solid lines) for the TE and TM resonant modes in racetrack resonators made of 4H-SiC from Cree (production grade). Note $\Delta\lambda_c$ for the TM mode has been multiplied by a factor of 10 for ease of comparison. All the measurements are carried out at a fixed modulation frequency of 10 kHz. (b) Same as (a) except that the 4H-SiC wafer (research grade) is from Shanghai Famous Trade ("SHFT" for short). (c) Frequency characterization for the TE$_{00}$ mode in (b): The blue stars are the measured and normalized (using $f=10$ kHz as the reference) modulation amplitude for a fixed $V_{pp}=5$ V. The red dashed line is the the simulated EO response using SiC's actual resistivity of \ $5 \times 10^{9}\ \Omega \cdot \text{cm}$, while the red solid line depicts the expected EO modulation amplitude in an optical microcavity with a linewidth of 280 MHz (corresponding loaded $Q$ of 0.7 million). The inset shows the recorded optical transmission corresponding to a modulation frequency of 1 GHz.}
\label{Fig_Cree}
\end{figure}

The frequency characterization in Fig.~\ref{Fig_freq}, in particular that the EO response above the MHz range is only limited by the cavity linewidth, provides strong evidence that the observed EO effect is the Pockels effect. To collaborate this result, we performed similar nanofabrication and EO measurements for 4H-SiC wafers from other wafer manufacturers \cite{Li_Kerr_compare}. Two such examples are provided in Fig.~\ref{Fig_Cree} for racetrack resonators fabricated on high-purity, semi-insulating 4H-SiC wafers from Cree \cite{Vuckovic_4HSiC_soliton} and Shanghai Famous Trade ("SHFT" for short) \cite{Vuckovic_4HSiC_MIcomb}. As can be seen, the extracted Pockels coefficients for these two materials are slightly different, but the overall characteristic, that the magnitude of $r_{33}$ is nearly zero while that of $r_{13}$ is on the order of $0.5$ pm/V, is consistent with the observation made from the Norstel wafer. In addition, the intrinsic $Q$s for microresonators made in these two wafers are lower (on the order of 1-2 million), mainly due to the slightly larger surface roughness from the polishing step. This allows us to check whether the EO modulation bandwidth can be proportionally enhanced by the increased cavity linewidth. The example shown in Fig.~\ref{Fig_Cree}(c) confirms our hypothesis, as decent EO modulation is observed up to 1 GHz for a cavity linewidth of 280 MHz and a $V_{pp}$ of 5 V. For convenience, we have summarized the estimated Pockels coefficients for all the 4H-SiC materials tested in this work in Table 1. 
 \begin{table}
\centering
\begin{tabular}{c c c c}
\hline
{\textbf{Pockels}} &\textbf{Norstel}&\textbf{SHFT}&\textbf{Cree}\\
\textbf{coefficient} &(pm/V)&(pm/V)&(pm/V)\\
\hline
$|r_{13}|$&$0.6\pm 0.1$&$0.4\pm 0.1$&$0.3\pm 0.1$\\
\hline
$|r_{33}|$&$\approx 0.02$&$\approx 0.01$&$\approx 0.01$ \\
\hline
\end{tabular}
\caption{Estimated magnitudes for the Pockels coefficients of 4H-SiC wafers from different manufacturers: Norstel is also known as ST Microelectronics; SHFT stands for Shanghai Famous Trade. All the measurements were carried out at a modulation frequency of 10 kHz.}
\label{Table1}
\end{table}

To conclude, we performed an extensive experimental investigation of the electro-optic effect in on-axis, semi-insulating 4H-SiC wafers from multiple manufacturers. Our results confirmed the existence of the Pockels effect and demonstrated GHz-level electro-optic modulation in 4H-SiC for the first time. Overall, the observed EO response is much stronger for the TE-polarized optical modes than the TM polarization (i.e., $|r_{13}|\gg |r_{33}|$). In addition, the EO effect is shown to be weakened at low modulation frequencies ($<1$ kHz), which is primarily attributed to the smaller resistivity of the 4H-SiC layer compared to that of the surrounding oxide. Nevertheless, our work points to a promising path towards achieving high-speed (>GHz) modulation and switching in the 4H-SiCOI platform, which is of critical importance for a wealth of chip-scale applications. 

\begin{backmatter}
\bmsection{Funding}
This work was supported by DARPA (D19AP00033) and NSF (2127499). 

\bmsection{Acknowledgments}
The authors would like to thank insightful discussions with Prof.~James A.~Bain and the equipment support from Prof.~Xu Zhang's group at ECE. J. Li also acknowledges the support of Axel Berny Graduate Fellowship from CMU. 

\bmsection{Disclosures}  The authors declare no conflicts of interest.

\bmsection{Data Availability} Data underlying the results presented in this paper are not publicly available at this time but may be obtained from the authors upon reasonable request.

\end{backmatter}
\bibliography{SiC_Ref}
\end{document}